\documentclass{ws-ijmpd}
\usepackage[super,compress]{cite}
\begin{document}

\markboth{Herdeiro and Radu}
{Spinning boson stars and hairy black holes with non-minimal coupling}

%
\catchline{}{}{}{}{}
%

\title{SPINNING BOSON STARS AND HAIRY BLACK HOLES 
\\
WITH NON-MINIMAL COUPLING}

\author{Carlos A. R. Herdeiro and Eugen Radu}

\address{Departamento de F\'\i sica da Universidade de Aveiro and CIDMA, \\ Campus de Santiago, 3810-183, Aveiro, Portugal\\ 
herdeiro@ua.pt, eugen.radu@ua.pt}

\maketitle

\begin{history}
\received{Day Month Year}
\revised{Day Month Year}
\end{history}

\begin{abstract}
We obtain spinning boson star solutions and hairy black holes with synchronised hair in the Einstein-Klein-Gordon model, wherein the scalar field is massive, complex and with a non-minimal coupling to the Ricci scalar. The existence of these hairy black holes in this model provides yet another manifestation of the universality of the synchronisation mechanism to endow spinning black holes with hair. We study the variation of the physical properties of the boson stars and hairy black holes with the coupling parameter between the scalar field and the curvature, showing that they are, qualitatively, identical to those in the minimally coupled case. By discussing the  conformal transformation to the Einstein frame, we argue that the solutions herein provide  new rotating boson star and hairy black hole solutions in the minimally coupled theory, with a particular potential, and that no spherically symmetric hairy black hole solutions exist in the non-minimally coupled theory, under a condition of conformal regularity.
\end{abstract}

\keywords{black holes; scalar fields; no-hair theorems}

\ccode{PACS numbers:}

\tableofcontents

\newpage

\section{Introduction}
Scalar fields play an important role in modern physics, entering a multitude of models
ranging from microscopic to cosmic scales.
Even though scalar fields are often used as a proxy to model more complex interactions,
the detection of the Higgs field shows the existence of fundamental
scalar fields  \cite{Aad:2012tfa}.

The interaction of scalar fields with gravity is a subject of long-standing interest.
Scalar fields are ubiquitous in modelling the physics of the early universe;
but there is also an intriguing possibility that such fields could play a role in black hole (BH) physics
\cite{Berti:2015itd,Herdeiro:2015waa}
or even cluster to form smooth horizonless compact objects: the so-called {\it boson stars} (BSs).\cite{Schunck:2003kk}
BSs are held together by their self-generated gravitational field,
and are supported against collapse  
by the dispersive effect due to the wave-like character of the scalar field \cite{Kaup:1968zz}.
They have been extensively studied in the literature,
being considered as possible dynamical astrophysical objects,\cite{Liebling:2012fv} 
black hole mimickers\cite{Vincent:2015xta,Cardoso:2016rao,Cao:2016zbh}
 or even as a possible constituent of dark matter.\cite{Suarez:2013iw,Hui:2016ltb}
Remarkably, spinning (but not static) BSs possess BH generalizations
 provided the scalar field co-rotates with the horizon: \textit{BHs with synchronised hair}.\cite{Herdeiro:2014goa,Herdeiro:2015gia}

It is therefore worthwhile studying more general couplings (than minimal coupling) of scalar fields to gravity.
Restricting to a single complex scalar field  $\Psi$,
perhaps the simplest and best motivated extension  is the inclusion
 of an explicit  coupling  between  $\Psi$  and
the Ricci scalar curvature of the spacetime, ${R}$, of the form
 $\xi\Psi^{*}\Psi {R} $,
 where $\xi$ is a dimensionless  coupling  constant and $\Psi^{*}$ denotes the complex conjugate field.
There are reasons to believe that such a nonminimal coupling term (as well as other couplings) appears naturally\cite{Capozziello:2011et,Faraoni:2000gx,Faraoni:1998qx}. 
For example, a nonminimal coupling is generated by quantum corrections even if it is absent
in the classical action and it is required in order to renormalize the theory \cite{BD}.
At a fundamental level, it is not known if there is a  preferential  value of $\xi$; however,
two cases occur most frequently in the
literature: ``minimal coupling" ($\xi$=0)  and ``conformal coupling"
($\xi$=1/6). The latter name is justified by the fact that conformal  invariance of the model  dictates $\xi$=1/6 for a massless
scalar field \cite{BD}.

In many  physical  situations,  the inclusion  of a $\xi\ne0$ term  leads to new
interesting  physical  effects even at the classical level.
Well known examples include 
the inflationary scenario with a nonminimally coupled ``inflaton" field\cite{Futamase:1987ua}
and
the Bronnikov-Melnikov-Bacharova-Bekenstein (BMBB) BH with conformal scalar hair\footnote{We remark that  the
BMBM BH is rather special.
The
spherically symmetric asymptotically flat BHs 
with generic $\xi$
cannot support nonminimally coupled spatially regular
neutral scalar fields 
\cite{Mayo:1996mv, Bekenstein:1996pn,Hod:2017ssh,Hod:2017hvl}.
}
\cite{Bekenstein:1974sf,BBM}.
More recent studies have considered
$e.g.$
traversible wormholes
\cite{Barcelo:1999hq,Barcelo:2000zf}, 
solitons and BHs with a non-minimally coupled gauged Higgs field,
\cite{vanderBij:2000cu,Nguyen:1993ep,Brihaye:2014vba} as discussed elsewhere.\cite{Faraoni:1998qx,Capozziello:2011et}

In this context, it is worthwhile investigating how a non-minimal coupling
affects the properties of  BHs with synchronised scalar hair together with their solitonic limit
-- spinning BSs. 
So far, only the spherically symmetric 
limit has been investigated
  \cite{vanderBij:1987gi,Horvat:2012aq,Horvat:2013plm},
in which case only solitonic solutions are possible.
In this paper we report spinning BHs and BH
solutions with synchronised hair of the Einstein-(complex, massive) Klein-Gordon equations with a nonminimal coupling  
$\xi\Psi^{*}\Psi {R} $, extending previous results\cite{Yoshida:1997qf,Herdeiro:2014goa} 
to this case.

\section{The framework}

\subsection{The  action and field equations}
We shall be working with an Einstein-Klein-Gordon (EKG) field theory, describing a massive complex scalar field 
$\Psi$ with a non-minimally coupling to Einstein's gravity, 
\begin{eqnarray} 
\label{action}
S=\int  d^4x \sqrt{-g}\left[ \frac{1}{16\pi G}R
   -\frac{1}{2} g^{ab}\left( \Psi_{, \, a}^* \Psi_{, \, b} 
	+ \Psi _{, \, b}^* \Psi _{, \, a} \right) 
 - \mu^2 \Psi^*\Psi
	- \xi \Psi^*\Psi R
 \right] ,
\end{eqnarray}
where $G$ is Newton's constant, 
 $\mu$ is the scalar field's mass 
and $\xi$ is a dimensionless constant.

The equations of the model are
\begin{eqnarray}
\label{E-eq}
&&
 R_{ab}-\frac{1}{2}g_{ab}R=8 \pi G~T_{ab}^{(eff)} \ ,
\\
&&
\label{KG-eq}
\nabla^2 \Psi=(\mu^2+\xi R)\Psi \ ,
\end{eqnarray}  
where we define the \textit{effective} energy-momentum tensor,
\begin{eqnarray}
&&
T_{ab}^{(eff)}=T_{ab}^{(\psi)}+T_{ab}^{(\xi)} \ ,
\end{eqnarray} 
which is the sum of two different contributions
\begin{eqnarray}
&&
T_{ab}^{(\psi)}=
 \Psi_{ , a}^*\Psi_{,b}
+\Psi_{,b}^*\Psi_{,a} 
-g_{ab}  \left[ \frac{1}{2} g^{cd} 
 ( \Psi_{,c}^*\Psi_{,d}+
\Psi_{,d}^*\Psi_{,c} )+\mu^2 \Psi^*\Psi\right]  ,
 \\
&&
 T_{ab}^{(\xi)}=2 \xi
\left[
\left(R_{ab}-\frac{1}{2}g_{ab}R\right)\Psi^*\Psi
+g_{ab}\nabla^2 (\Psi^*\Psi)
-(\Psi^*\Psi)_{;ab}
\right] \ .
\end{eqnarray}  
%

It is useful to define an
``effective" Newton's constant
\begin{eqnarray} 
\label{Geff}
\frac{1}{G_{eff}}\equiv \frac{1}{G}-32 \pi \xi \Psi^*\Psi .
\end{eqnarray}

\subsection{A Smarr relation}
\label{sec_Smarr}
We are interested in stationary, axisymmetric, asymptotically flat BH spacetimes. These geometries possess: 1) a unique
translational Killing vector $K^a$ which is timelike and normalized to $K^a K_a=-1$
near spatial infinity; 2)  a unique rotational Killing vector $\tilde K^a$ normalized
by demanding that its orbits to be closed curves with parameter length $2\pi$.
Then, a  null
vector $\chi$ exists that is tangent to the horizon  null generator; it can be
expressed as
\begin{eqnarray} 
\chi=K+\Omega_H \tilde K~,
\end{eqnarray}
where
$\Omega_H$
is the angular velocity of the BH.
In general both $K$ and $\tilde K$ act nontrivially on the scalar $\Psi$,
 and only the energy-momentum tensor remains invariant
(in particular $K \Psi^*\Psi= \tilde K \Psi^*\Psi=0$).

Following a well known procedure\cite{Bardeen:1973gs,Townsend:1997ku}, the energy and angular momentum content of the spacetime may be split into different components. One starts with the identities
\begin{eqnarray} 
\label{BCH}
M=M_H+M_{(\xi)}+M_{(\Psi)}, \qquad J=J_H+J_{(\xi)}+J_{(\Psi)},
\end{eqnarray}
where 
$M$ and $J$ are the ADM mass and total angular momentum,
and
\begin{eqnarray} 
\label{h1}
M_H=-\frac{1}{4\pi G}\oint _H dS_{ab} \nabla^a K^b, \qquad 
J_H=\frac{1}{16\pi G}\oint _H dS_{ab} \nabla^a \tilde K^b \ ,
\end{eqnarray}
are the horizon mass and angular momentum, computed as Komar integrals on a spatial section of the horizon.  The matter contributions are given by 3-volume integrals that read
\begin{eqnarray} 
M_{(\xi),(\Psi)}=-2\int _\Sigma dS_{a}  \left( T_b^{(\xi),(\Psi)a} K^b-\frac{1}{2}T^{(\xi),(\Psi)}K^a \right), \nonumber \\ 
J_{(\xi),(\Psi)}=\int _\Sigma dS_{a}  \left( T_b^{(\xi),(\Psi)a} \tilde K^b-\frac{1}{2}T^{(\xi),(\Psi)}\tilde K^a \right).
\label{rel1}
\end{eqnarray}
Remarkably, for regular configurations 
 with an  asymptotically vanishing scalar field,  
a straightforward computation shows that
the  non-minimal coupling contribution to both $M$ and $J$ can be expressed as a
total divergence, the only non-vanishing contribution being the  horizon surface integral,
\begin{eqnarray} 
\label{h2}
M_{(\xi)}=2\xi \oint _H dS_{ab} \Psi^*\Psi \nabla^a K^b, \qquad 
J_{(\xi)}=- \xi \oint _H dS_{ab} \Psi^*\Psi \nabla^a \tilde K^b~.
\end{eqnarray}
As usual \cite{Bardeen:1973gs}, one can express $dS_{ab}$ as $\chi_{[a} n_{b]} dA$,
where $n_a$ is the other null vector field orthogonal to the horizon,
normalized as $n_a \chi^a=-1$, and $dA$ is the surface area element of the horizon.
After replacing relations (\ref{h1}) and (\ref{h2}) into (\ref{BCH}),
one arrives at the following Smarr mass formula which relates 
the temperature, entropy and the global charges
\begin{eqnarray}
\label{Smarr}
M=2 T_H S +2\Omega_H (J-J_{(\Psi)})+ M_{(\Psi)}.
\end{eqnarray}
Here, $T_H=\kappa/2\pi$ is the Hawking temperature of the  BHs  (where $\kappa$ is the surface gravity, defined as usual in terms of $\chi$),
while $M_{(\Psi)}$,
and $J_{(\Psi)}$,
are the scalar field energy and angular momentum $outside$ the BH, 
 given by (\ref{rel1}).
 
The entropy $S$ of the BHs has an extra contribution with respect to that in Einstein's gravity, 
due to the non-minimal coupling with the scalar field
\begin{eqnarray}
\label{entropy0}
S=S_E+S_{(\xi)}, \qquad {\rm with}~~
S_E=\frac{1}{4G}\int_{H} dA, \qquad 
S_{(\xi)}=- 4\pi \xi \int_{H}  \Psi^*\Psi d A,~
\end{eqnarray}
a relation which can also be written in the suggestive form\footnote{The same expression of entropy
is derived by using Wald's formalism\cite{Wald:1993nt,Iyer:1994ys}.
}
\begin{eqnarray}
\label{entropy}
S=\frac{1}{4}\int_{H} \frac{1}{G_{eff}} d A~,
\end{eqnarray}
where $G_{eff}$
is the effective Newton's constant given by 
(\ref{Geff}).

The solutions should also satisfy the first law of BH thermodynamics:
\begin{eqnarray}
\label{first-law}
dM=T_H dS +\Omega_H dJ .
\end{eqnarray}

\subsection{The Ansatz and quantities of interest}

The non-minimally coupled BSs and hairy BHs (HBHs) are constructed with the same ansatz used in previous works.
\cite{Herdeiro:2014goa,Herdeiro:2015gia}
Working  in a coordinate system with $K=\partial_t$ and
$\tilde K=\partial_\varphi$,
we consider  a line element
\begin{eqnarray}
\label{ansatz}
ds^2=e^{2F_1}\left(\frac{dr^2}{N }+r^2 d\theta^2\right)+e^{2F_2}r^2 \sin^2\theta (d\varphi-W dt)^2-e^{2F_0} N dt^2, \\ 
~~{\rm with}~~N\equiv 1-\frac{r_H}{r}\ ,
\end{eqnarray} 
and a scalar field
\begin{eqnarray}
\Psi=\phi(r,\theta)e^{i(m\varphi-w t)}~.
\label{scalar_ansatz}
\end{eqnarray} 
Consequently, the problem contains five unknown functions $\{F_i,W;\phi\}$
which depend on $(r,\theta)$ only.
Also, $w$ is the scalar field frequency and $m=\pm 1,\pm 2$\dots is the azimuthal harmonic index.
 Without loss of generality, we take $w>0$.

The BHs have a horizon located at $r=r_H$; the range of $r$ considered in numerics is $r_H\leqslant r<\infty$.
 BSs correspond to the $r_H=0$ limit of (\ref{ansatz}).
 Most of the quantities of interest are 
encoded in the expression for the metric functions at the horizon or at infinity.
Considering first horizon quantities,  the 
Hawking temperature $T_H$, the event horizon area $A_H$ and the event horizon velocity $\Omega_H$
are 
\begin{eqnarray}
\label{THAH}
T_H=\frac{1}{4\pi r_H}e^{(F_0 -F_1)|_{r_H} } \ ,
\qquad 
A_H=2\pi r_H^2 \int_0^\pi d\theta \sin \theta~e^{(F_1 + F_2)|_{r_H} } \ , \\
\Omega_H=-\frac{g_{\varphi t}}{g_{tt}}\bigg|_{r_H}=W \bigg|_{r_H}~.
\end{eqnarray}  
The ADM mass $M$ and the angular momentum $J$ are read from 
the asymptotic sub-leading behaviour of the metric functions:
\begin{eqnarray}
\label{asym}
g_{tt} =-e^{2F_0}N+e^{2F_2}W^2r^2 \sin^2 \theta
=-1+\frac{2GM}{r}+\dots,
\\
g_{\varphi t}=-e^{2F_2}W r^2 \sin^2 \theta=-\frac{2GJ}{r}\sin^2\theta+\nonumber \dots. \\
\end{eqnarray}  

One notices also that
the action (\ref{action}) is invariant under the global $U(1)$ transformation $\Psi\rightarrow e^{i\alpha}\Psi$, 
where $\alpha$ is constant. Thus, the scalar 4-current, $j^a=-i (\Psi^* \partial^a \Psi-\Psi \partial^a \Psi^*)$, is conserved: 
 $j^a_{\ ;a}=0$. 
It follows that integrating the timelike component of this 4-current in a spacelike slice $\Sigma$ yields a conserved quantity -- the \textit{Noether charge}:
\begin{eqnarray}
\label{Q}
Q=\int_{\Sigma}~j^t d\Sigma_t \ ,
\end{eqnarray}
with the explicit expression
\begin{eqnarray}
\label{Q-int}
Q=2\pi \int_{r_H}^\infty dr \int_0^\pi d\theta  
~r^2\sin \theta ~e^{F_0+2F_1+F_2}  \frac{m(w-mW)}{N}\phi^2 .
\end{eqnarray}
The relation expressing the quantisation of the total angular momentum\cite{Herdeiro:2015gia}
\begin{eqnarray}
\label{QJ}
J_{(\Psi)}=m  Q
\end{eqnarray}
holds for arbitrary values of $\xi$
and can be used to further simplify 
the Smarr relation
(\ref{Smarr}).

\section{Non-minimally coupled boson stars and BHs with scalar hair}

\subsection{Boundary conditions}

The solutions with $\xi\neq 0$
are constructed by using the same approach as in the minimally coupled case, described at length in our previous work.\cite{Herdeiro:2015gia}
At spatial infinity, asymptotic flatness dictates
\begin{equation}
\lim_{r\rightarrow \infty}{F_i}=\lim_{r\rightarrow \infty}{W}=\lim_{r\rightarrow \infty}{\phi}=0 \ ,
\end{equation}
while on the symmetry axis $\theta=0,\pi$ we impose
\begin{equation}
\partial_\theta F_i = \partial_\theta W = \phi = 0.
\end{equation}
The numerical treatment of the problem is simplified by defining a new radial coordinate
$x=\sqrt{r^2-r_H^2}$, such that the horizon is located at $x=0$.
There we impose\footnote{Note that the scalar field does not vanish at the horizon, 
 being a function of $\theta$.}
\begin{equation}
\partial_x F_i \big|_{r=r_H}= \partial_x \phi  \big|_{r=r_H} =  0, \qquad W \big|_{r=r_H}=\frac{w}{m}.
\end{equation}
Note that the horizon boundary condition imposed on $W$
encodes the 
{\it 'synchronization condition'}  \cite{Herdeiro:2014goa}
\begin{eqnarray}
\label{cond}
w=m\Omega_H,
\end{eqnarray}
which implies 
that there is no flux of the scalar field into (or from) the BH, 
$\chi^{\mu}\partial_\mu \Psi=0$.

The BSs do not possess a horizon and therefore $0\leqslant r<\infty$.
At the origin ($r=0$) we impose
\begin{equation}
\partial_r F_i = \partial_r W = \phi = 0.
\end{equation}

For any $\xi$, an approximate form of the solution can be constructed on the boundary of the domain of integration,
which is compatible with the boundary conditions given above.
Also, all solutions reported in this work are symmetric $w.r.t.$
a reflection on the equatorial plane $\theta=\pi/2$.
Moreover,  we focus on nodeless solutions 
(with no zeros of the scalar field in the equatorial plane)
since these are typically the most stable ones.

The numerical integration is performed with dimensionless variables
introduced
by using
natural units set by $\mu$ and $G$,
\begin{eqnarray}
r\to r/\mu,~~
\phi \to \phi /\sqrt{4 \pi G},~~
w \to w/\mu~.
 \end{eqnarray}
 As a result, no dependence on either $G$ or $\mu$ is present in the equations.

\subsection{Boson star solutions}

\begin{figure}[ht]
\begin{center}
\includegraphics[width=0.7\textwidth]{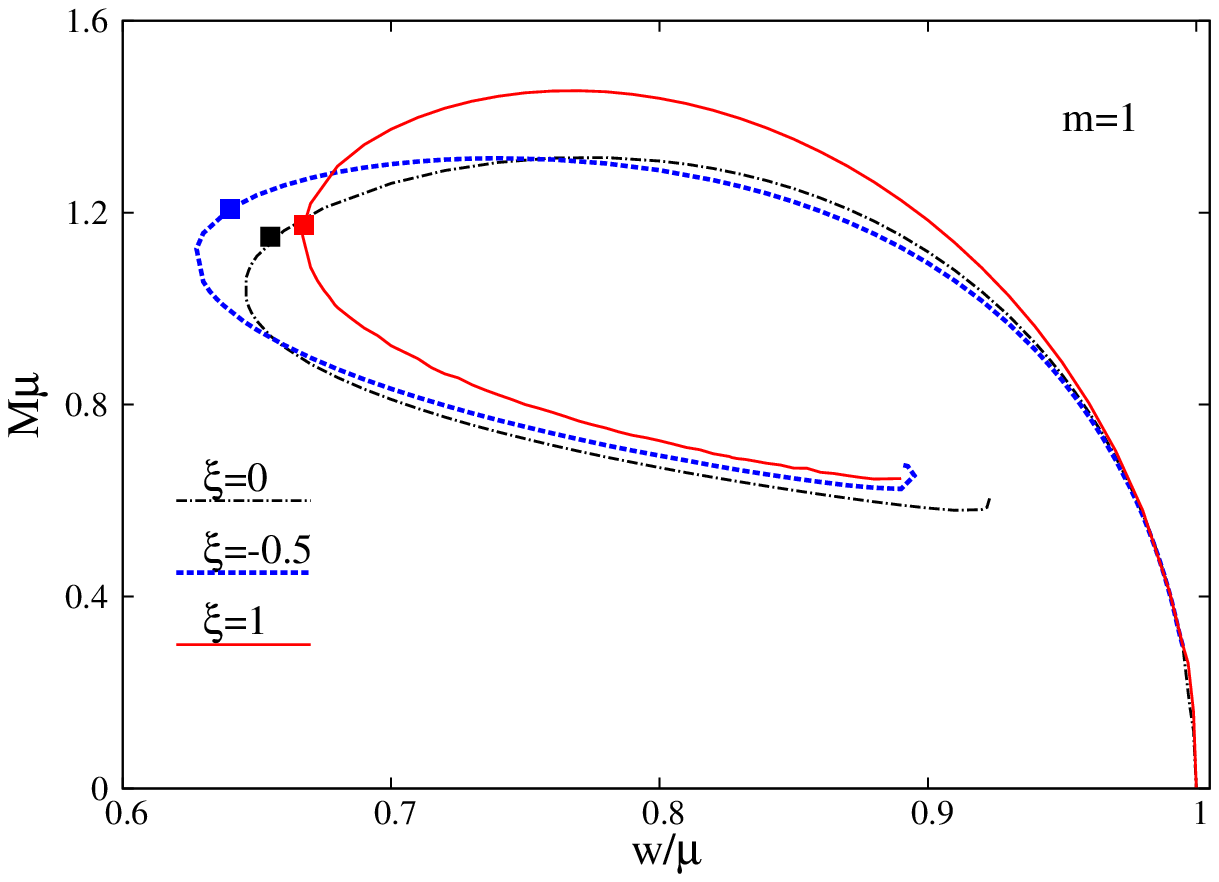}\\
\includegraphics[width=0.7\textwidth]{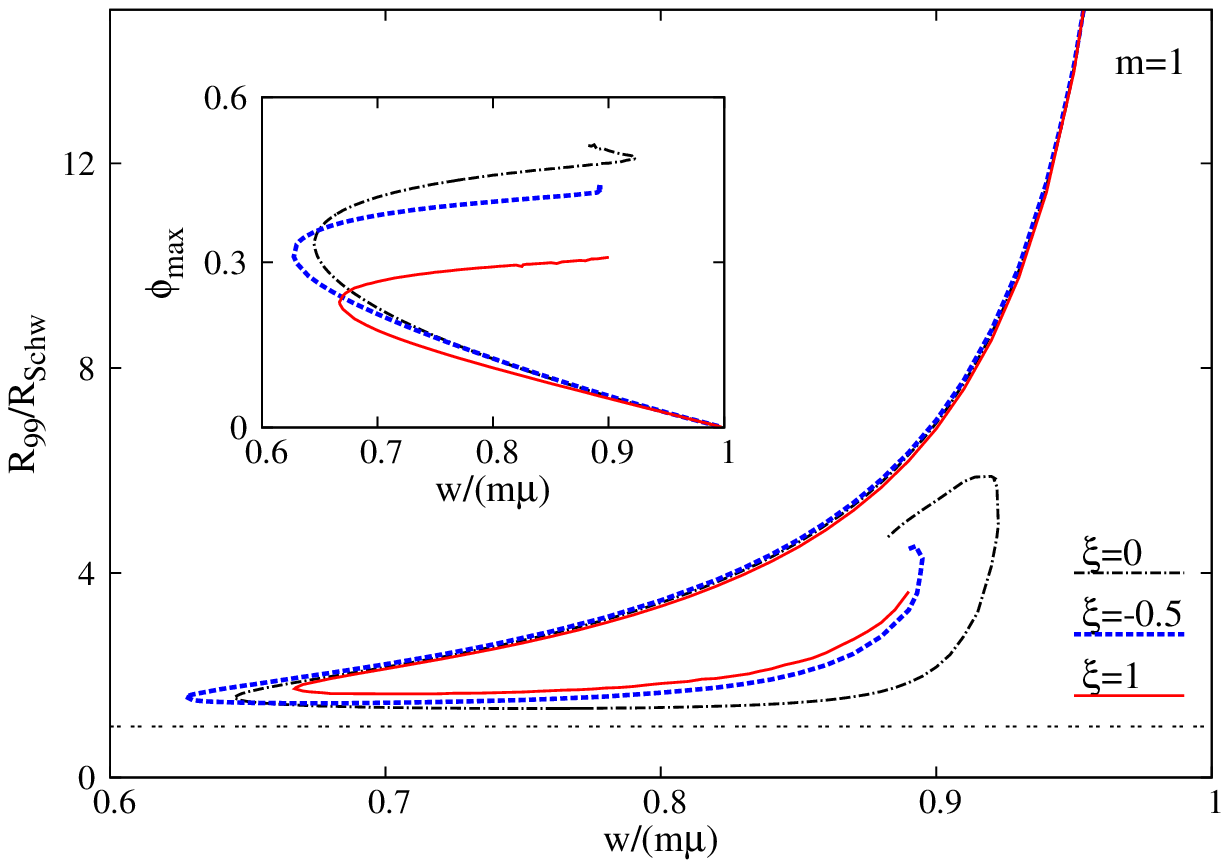}
\end{center}
\caption{\small (Top panel) The  frequency-mass  diagram is shown for families of spinning boson stars with $m=1$ and three different values of
the coupling constant $\xi$.
The points highlighted with a square indicate the appearance of an ergo-region (an ergo-torus) when moving towards the centre of the spiral. (Bottom panel) 
Inverse compactness of BSs is shown for the same solutions.
 The inset shows the maximal value of the scalar field
along the BS lines.  }
\label{BSwM} 
\end{figure}

\begin{figure}[ht]
\begin{center}
\includegraphics[width=0.7\textwidth]{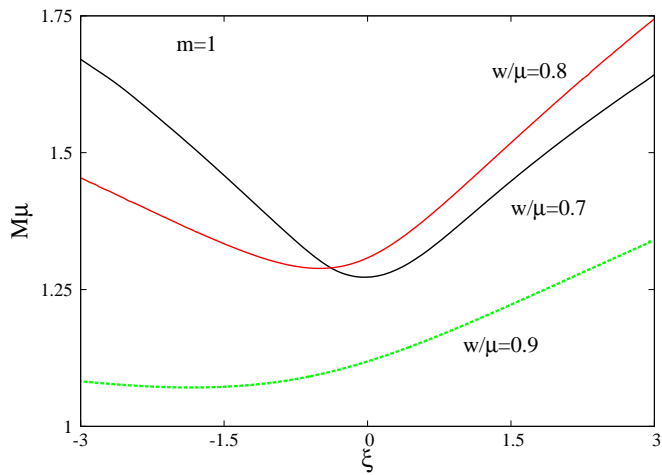}
\end{center}
\caption{\small The BSs mass is shown as a function of $\xi$ 
for three different frequencies. }
 \label{xi}
\end{figure}

The minimally coupled spinning 
BSs ($\xi=0$)
were originally studied, independently, by Schunck and Mielke\cite{Schunck:1996he} and Yoshida and Eriguchi\cite{Yoshida:1997qf}. A relevant more recent work is due to Grandclement, Som and Gourgoulhon.\cite{Grandclement:2014msa}. Turning on the non-minimal coupling, some results of the numerical integration are exhibited in Figs. \ref{BSwM} and \ref{xi}.

 %
As one can observe from Fig. \ref{BSwM} (top panel), the basic features of the mass-frequency diagram found for the minimally coupled solutions are preserved.
Firstly, for any coupling,
the  BSs exist for $w<\mu$; this is a bound state condition
which emerges from the long range behavior of the scalar field, where the Ricci scalar tends to zero.
As we decrease the frequency, 
the mass increases until a maximum value\footnote{In the spherical case, this maximal value is proportional\cite{vanderBij:1987gi} to $\sqrt{|\xi|}$.
For spinning solutions, however, the numerical accuracy deteriorates before 
large enough values of $|\xi|$ can be considered; thus a similar relation could not be derived.}, which is always
of the order of $1/\mu$.
Further decreasing $w$ one finds a minimal frequency 
$w_{min}$,
 below which no BS solutions are found.
This minimal frequency 
increases as $\xi$ increases, and
decreases for negative values of the coupling constant.
 Then, for any $\xi$, the BS curve seems to spiral towards a central region of the diagram
 where numerical accuracy deteriorates.
 Qualitatively, this is also the behaviour
found for spherically symmetric BSs ($m=0$), 
in which case, 
a detailed investigation of the inspiraling behaviour was possible.
We have found a similar behaviour for the total angular momentum, $J$, or equivalently, $cf.$~\eqref{QJ}, for the Noether charge.

As a particular physical property, for any $\xi$, a part of fast spinning BSs possess a toroidal ergo-surface, likewise the minimally coupled case.\cite{Herdeiro:2014jaa,Herdeiro:2016gxs}
In a frequency-mass diagram, this corresponds to the inner part of the spiral 
starting with a critical configuration marked with a square in Fig. \ref{BSwM} (top panel).

As another physical property, one may wonder how 
a nonzero $\xi$ affects the compactness of the spinning BSs.
Following previous literature\cite{AmaroSeoane:2010qx}
we define the inverse compactness by comparing $R_{99}$, defined as the circumferential radius wherein 99\% of the BS's mass is contained,  with the Schwarzschild radius associated to that mass, 
$R_{Schw}=2M_{99}$:
\begin{equation}
{\rm Compactness}^{-1}\equiv  \frac{R_{99}}{2M_{99}} \ .
\label{compactness}
\end{equation}
The result for the inverse compactness of BSs with several values of 
$\xi$ is exhibited in Figure~\ref{BSwM}
(bottom panel). 
One can see that the inverse compactness is always greater than unity; in other words, BSs are less compact than BHs, as one would expect.
Also, at least for the considered values, the non-minimal coupling $\xi \neq 0$ does not alter substantially the compactness of the BSs, tending to make such compactness smaller, for fixed frequency, in the strongest gravity region we have considered (towards the centre of the spiral).

 Further insight on the influence of a nonzero coupling on the mass of the BS solutions
can be found in Fig. \ref{xi},
 where the coupling constant $\xi$
is varied for three fixed values of the frequency $w$.
Interestingly, while for a positive coupling, $M$ always increases with $\xi$ (for fixed frequency),
a minimal mass value is approached for some negative
$\xi$, with $M$ increasing with $|\xi|$ for lower $\xi$ values.

\subsection{Black holes with synchronized scalar hair}

Before discussing the non-minimally coupled HBHs, let us recall that 
for $\xi=0$, the emergence of scalar
hair can be seen in linearized theory, by considering the massive
Klein-Gordon equation 
$(\nabla^2-\mu^2 )\Psi=0$, as a test field, 
on a Kerr BH background. 
This was interpreted\cite{Herdeiro:2014goa}
 as a zero mode of the superradiant
instability \cite{Brito:2015oca}.
That is,
by solving the Klein-Gordon equation on the Kerr background, 
real frequency bound states can be obtained when $w=m\Omega_H$, corresponding to linearized (hence non-backreacting) hair, called \textit{stationary scalar clouds}~\cite{Hod:2012px,Hod:2013zza,Herdeiro:2014goa,Hod:2014baa,Benone:2014ssa}.
However, since the Kerr BH solves the vacuum Einstein equations $(R=0)$,
it is obvious that all results in the aforementioned works remain valid in the 
non-minimally coupled case.
Thus one can predict that HBHs with $\xi\neq 0$
will branch off from the same set of Kerr BHs -- forming 
 the {\it fundamental existence line}\cite{Herdeiro:2014goa} -- as the ones with $\xi=0$, 
a feature which is confirmed by our numerical solutions.
However, when deviating from that line, the Ricci scalar
deviates from zero and the effects of a non-zero coupling term
become  relevant.  

A different path in constructing HBHs is to start instead with their solitonic limit.
We have found that, for a given $\xi$, one can add a small BH at the center 
of any spinning BS, regardless of $w$. 
By increasing the horizon size from zero (via the parameter $r_H$), 
we obtain rotating BH solutions with $\Omega_H$
fixed by (\ref{cond}).
 
\begin{figure}[ht]
\begin{center}
\includegraphics[width=0.7\textwidth]{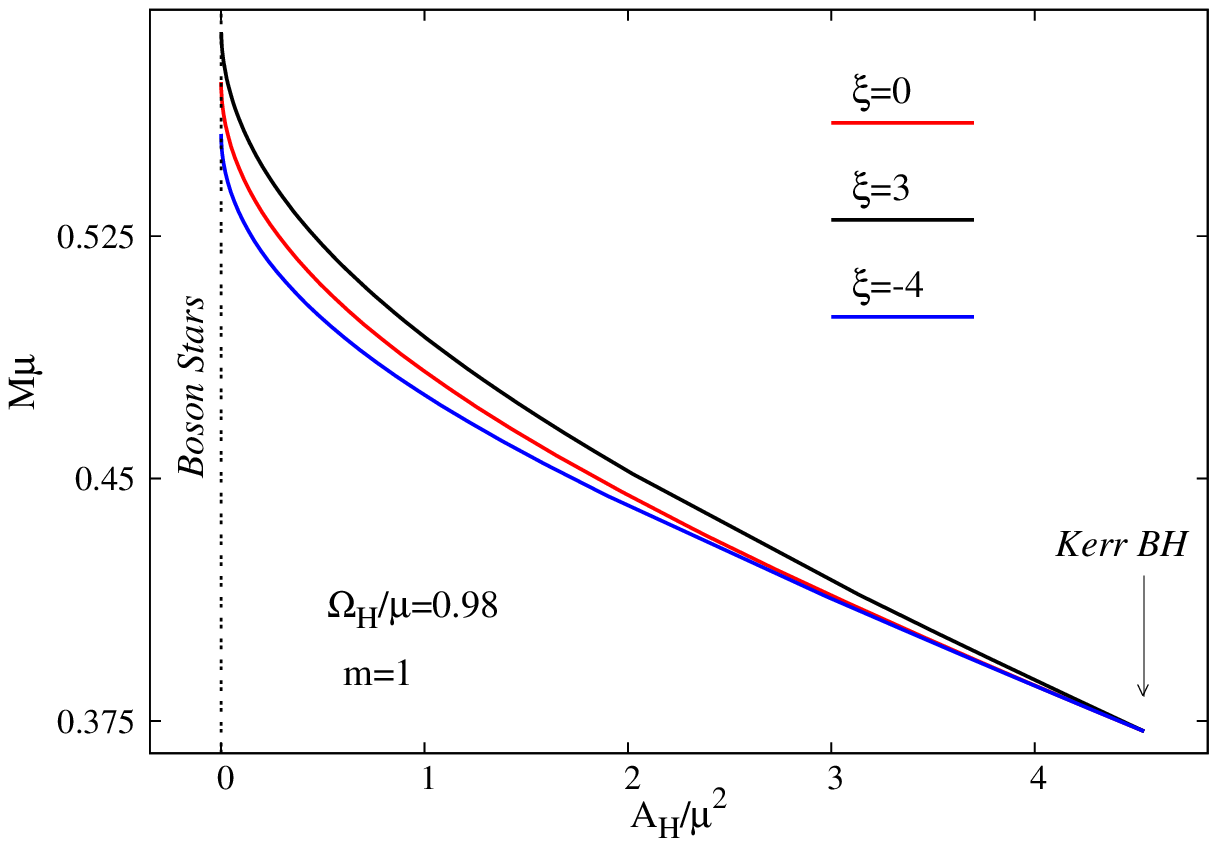}\\
\includegraphics[width=0.7\textwidth]{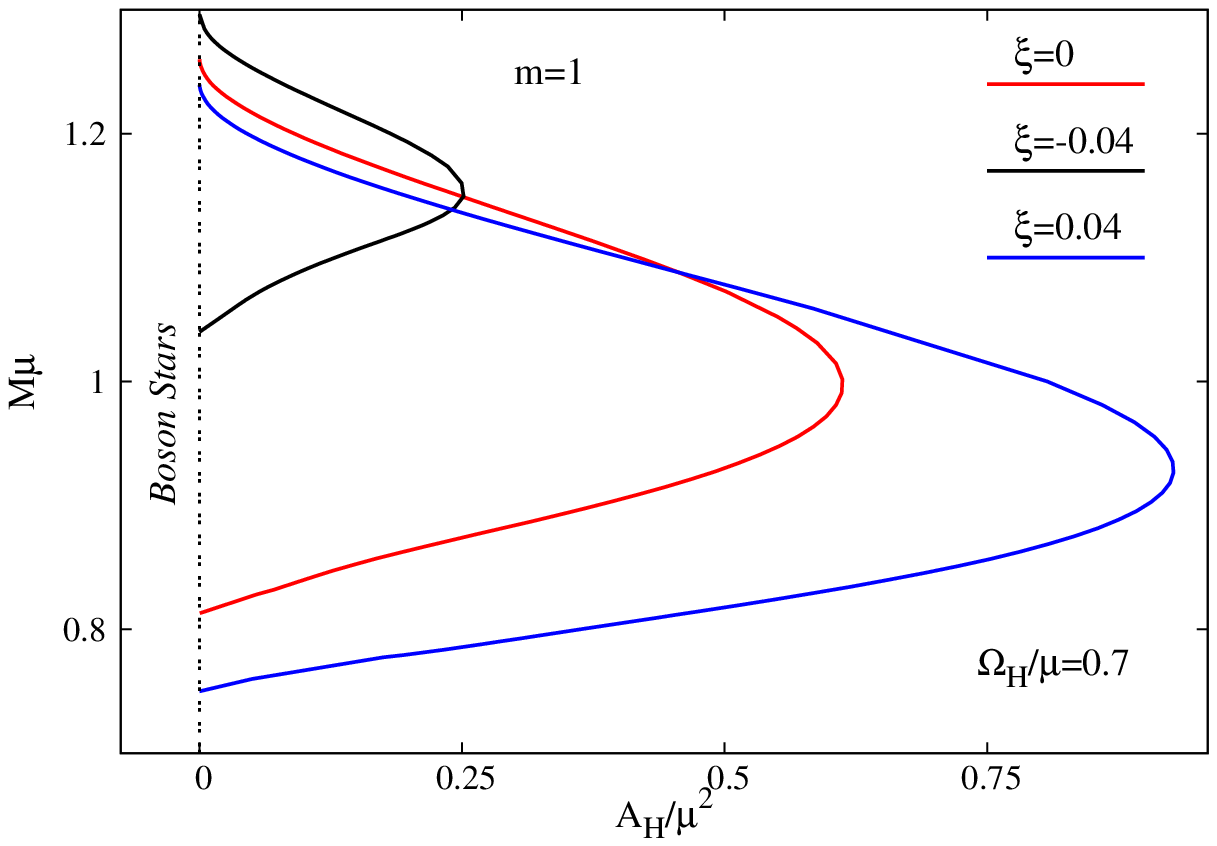}
\end{center}
\caption{\small Mass of the hairy BHs, in units of the scalar field mass, $vs.$ the horizon area
 for three sets of solutions with the same angular velocity 
and different values of the coupling constant $\xi$. }
\label{BH1}
\end{figure}

A systematic, thorough study of the non-minimally coupled HBH solutions is beyond the scope of this work.
Instead, in order to establish their existence and probe some basic properties, we have considered several values of  $\xi$
and a set of frequencies with $w_{min}<w<\mu$.
In each case, we have studied HBHs for all allowed range of the $r_H$.
The numerical results strongly suggest that 
all basic properties of the minimally coupled HBHs are kept.
 For any $\xi$, the region where  HBHs exist is delimited by: 
$(i)$ the BS curve  where the horizon shrinks to zero size, 
$(ii)$ the subset of Kerr solutions that support the fundamental existence line of stationary scalar clouds,
and 
$(iii)$ a set of  extremal ($i.e.$ zero temperature) HBHs.

Some results of the numerical integration are exhibited in Fig. \ref{BH1}
where we show the mass $M$ as a function of horizon area $A_H$
for several sets of HBHs solutions with fixed values of $\Omega_H$ (or, equivalently, $w$).

For a large enough frequency - $e.g.$ $w/m=0.98$ in Fig. \ref{BH1} (top panel) -,
the HBHs interpolates between the corresponding BSs and the vacuum Kerr solution
on the existence line.
As $r_H\to 0$ the horizon area vanishes while the temperature
 diverges.
For sufficiently small values\footnote{For some intermediate frequencies, one finds also branches of HBHs
with fixed $w$ staring in BSs and ending in extremal BHs with scalar hair. 
These limiting configurations have finite
horizon size and global charges.}  of $w/\mu$,
the solutions interpolate between two  
BS solutions with the same scalar field frequency  - $e.g.$ $w/m=0.7$ in Fig. \ref{BH1} (bottom panel).

Let us conclude with two final remarks. Firstly, for all solutions studied so far, 
the contribution of the non-minimal coupling term to the total entropy (\ref{entropy0}),
 was always several orders of magnitude smaller than the Einstein term. Secondly,  that
the bound for the horizon linear velocity $v_H$,\cite{Herdeiro:2015moa} $v_H<1$, 
is fulfilled for all HBH solutions considered so far.

\section{Further remarks}

In this paper we have constructed rotating boson star solutions and initiated the study of hairy BHs in the EKG model with a non-minimal coupling of the scalar field to the Ricci tensor. One of the conclusions of our study is that the synchronisation mechanism to endow spinning BHs with hair\cite{Herdeiro:2014ima} survives yet another generalisation: considering a non-minimal coupling between the scalar field and the Ricci scalar. This adds up to the already extensive list of examples where this synchronisation mechanism allows the construction of hairy BHs, including:  different matter fields (scalar\cite{Herdeiro:2014goa,Herdeiro:2015gia,Herdeiro:2015tia} and vector\cite{Herdeiro:2016tmi,Herdeiro:2017phl}), charged BHs,\cite{Delgado:2016jxq} in spacetime dimensions different than four\cite{Brihaye:2014nba,Herdeiro:2015kha}, for non-asymptotically flat spacetimes,\cite{Dias:2011at} with non-spherical horizon topologies\cite{Herdeiro:2017oyt} and in scalar-tensor theories.\cite{Kleihaus:2015iea}
 
 The physical picture concerning both the solitons and hairy BHs with non-minimal coupling is qualitatively similar to that obtained in the minimally coupled case. Two interesting technical aspects that we would like to emphasise are: 1) that the contribution of the non-minimal coupling to the ADM mass and angular momentum of a boson star, $ M_{(\xi)},J_{(\xi)}$, $cf.$~\eqref{rel1}, turns out to vanish globally, albeit being non-zero locally, $cf.$ eq.~\eqref{h2}; 2) That an elegant form for the entropy in terms of an effective Newton's constant can be obtained, $cf.$~\eqref{entropy}. The latter observation has been known in the literature\cite{Ashtekar:2003jh}; to the best of our knowledge, however, the transformation of the volume integrals~\eqref{rel1} for $ M_{(\xi)},J_{(\xi)}$ into the surface integrals~\eqref{h2}; has not been previously observed.

We would like to close by remarking that the solutions we have presented herein allow us to obtain, easily, a new set of solutions of the non-minimally coupled EKG model, with a particular choice of potential.  By using the conformal rescaling of the action (\ref{action}):
\begin{eqnarray} \label{transform1}
\bar{g}_{\mu\nu}=\Omega^{2} g_{\mu\nu},
\quad
\Omega^{2}=1-16 \pi G \xi \Psi^*\Psi=\frac{G}{G_{eff}} ,
\end{eqnarray}
the model we have considered can be mapped to the Einstein frame, that is a minimally coupled scalar field theory to gravity. 
$\Omega^{2}>0$ has a clear physical  meaning since it implies $G_{eff}>0$,
a condition which is satisfied by all solutions reported in this work.  
The pairs of variables  
(metric $g_{ab}$ and scalar $\Psi$)
defined  originally constitute what is called
a Jordan frame. 
 Consider now the transformation 
\begin{eqnarray}
 \label{transform2}
d\bar \Psi=\frac{\sqrt{1-16\pi G \xi(1-6\xi)  \Psi^*\Psi}}{1-16 \pi G \xi \Psi^*\Psi}d\Psi,
\end{eqnarray}
such  that, in the redefined  action
\begin{eqnarray} 
\label{newaction}
S=\int d^{4}x\sqrt{-\bar{g}}\left[\frac{\bar{R}}{16\pi G}
   -\frac{1}{2} \bar g^{ab}\left( \bar \Psi_{, \, a}^*  \bar \Psi_{, \, b} 
	+  \bar \Psi _{, \, b}^* \bar \Psi _{, \, a} \right) 
- V(\bar \Psi)\right] ,
\end{eqnarray}
the scalar field $\bar \psi$ becomes  minimally  coupled to $\bar R$.
One pays the price, however, that the new scalar potential becomes more complicated, 
\begin{equation}
V(\bar \Psi)=\frac{\mu^2 \Psi^*\Psi}{(1-16 \pi G \xi \Psi^*\Psi)^2}~,
\end{equation}
with $\Psi$ a function of $\bar \Psi$, via the transformation (\ref{transform2}).
  
The new variables (metric $\bar{g}_{ab}$ and scalar $\bar \Psi$)
are said to constitute an Einstein  frame.
 The transformation 
given by eqs. (\ref{transform1}, \ref{transform2}) therefore maps a solution of the field 
equations 
(\ref{E-eq}),
(\ref{KG-eq})
 to a solution 
that extremizes (\ref{newaction}). The transformation is independent of any assumption of 
symmetry and, in this sense, it is covariant; one can easily infer that the 
transformation is one-to-one in general.

Therefore, all spinning solutions of the initial model  (\ref{action})
are mapped to spinning  BSs and HBHs of the Einstein  frame model (\ref{newaction}).
We note that the mass, angular momentum, Hawking temperature, horizon angular velocity 
together with the synchronization condition 
are not 
affected by the transformation  (\ref{transform1}), (\ref{transform2}) (we recall that $\Psi \to 0$ asymptotically), while
the entropy together with the mass and angular momentum stored 
in the scalar field are different in the two frames.
 
Furthermore,
the Weyl rescaling (\ref{transform1}) helps us to rule out the existence of static, spherically symmetric
BH solutions in the original model 
(\ref{action}) as long as $\Omega^2>0$.
This can be shown as follows.
We  start by supposing the  
existence of a static BH  solution in the original Jordan frame.
Since the transformation
(\ref{transform1}, \ref{transform2}) 
 preserves symmetries and the structure of the light cone,
this results in a BH solution in the  Einstein  frame.
However, this contradicts a well known no-hair theorem\cite{Pena:1997cy}, that applies for a scalar field minimally coupled with gravity.
This theorem applies to scalar fields that may (but need not to) vary harmonically with time and may have an arbitrary positive semidefinite potential.
One can easily show that the proof of this no-hair theorem still applies
for the model (\ref{newaction}),
since, in particular, the potential
$V(\bar \psi)$ is positive semidefinite.
Therefore we conclude that no  BH  solutions exist also in the original Jordan frame. The existence of the BBMB solution circumvents this argument since $\Omega^2$ is not positive everywhere outside the horizon. Thus, this solution violates a requirement we may call  \textit{conformal regularity}: that the conformal factor $\Omega^2$ is non negative outside the horizon.

\section*{Acknowledgments}

C. H. and E. R. acknowledge funding from the FCT-IF programme.  This project has received funding
from  the  European  Union's  Horizon  2020  research  and  innovation  programme  under  the H2020-MSCA-RISE-2015 Grant No.   StronGrHEP-690904, the H2020-MSCA-RISE-2017 Grant No. FunFiCO-777740  and  by  the  CIDMA  project UID/MAT/04106/2013. 
The authors  would also  like  to  acknowledge networking support by the COST Action GWverse CA16104.



\end{document}